\documentclass[11pt]{article}

\usepackage{graphicx}
\usepackage{epstopdf}
\usepackage{amsmath}
\usepackage{amssymb}
\usepackage{amsfonts}
\usepackage{amsthm}
\usepackage[usenames]{color}
\usepackage{array}

\usepackage[
      colorlinks=true,
      linkcolor=blue,
      urlcolor=blue,
      filecolor=blue,
      citecolor=red,
      pdfstartview=FitV,
      pdftitle={},
      pdfauthor={},
      pdfsubject={},
      pdfkeywords={},
      pdfpagemode=None,
      bookmarksopen=true
]{hyperref}

\usepackage{epsfig}

\usepackage{hyperref}

\newcommand{\bea}{\begin{eqnarray}}
\newcommand{\eea}{\end{eqnarray}}
\newcommand{\ba}{\begin{array}}
\newcommand{\ea}{\end{array}}
\newcommand{\ee}{\end{equation}}

\numberwithin{equation}{section}

\begin{document}

\begin{flushright}
\texttt{\today}
\end{flushright}

\begin{centering}

\vspace{2cm}

\textbf{\Large{
Holographic Calculation of BMSFT Mutual and \\ 3-partite Information  }}

  \vspace{0.8cm}

  {\large   Mohammad Asadi, Reza Fareghbal }

  \vspace{0.5cm}

\begin{minipage}{.9\textwidth}\small
\begin{center}

{\it  Department of Physics, 
Shahid Beheshti University, 
G.C., Evin, Tehran 19839, Iran.  }\\

  \vspace{0.5cm}
{\tt  m$\_$asadi@sbu.ac.ir,  r$\_$fareghbal@sbu.ac.ir}
\\ $ \, $ \\

\end{center}
\end{minipage}


\begin{abstract}
We use  flat-space holography to calculate  the mutual information and the 3-partite information of a two-dimensional BMS-invariant field theory (BMSFT$_2$). This theory  is the putative holographic dual of the three-dimensional asymptotically flat spacetimes. We find a bound   in which  entangling transition occurs for  zero and  finite temperature BMSFTs.  We  also show that the  holographic 3-partite information is always non-positive which indicates that  the holographic mutual information  is monogamous.
\end{abstract}

\end{centering}

\newpage



\section{Introduction}
The gauge/gravity duality provides a remarkable framework to study  key features of the boundary field theory  dual to some gravitational theory on the bulk side. The famous example of gauge/gravity duality is  the AdS/CFT correspondence which proposes a duality between asymptotically AdS spacetimes in (d+1)-dimensions and d-dimensional conformal field theories \cite{Maldacena}.

 Asymptotically AdS (AAdS) geometries are solutions of Einstein gravity with negative cosmological constant whose AdS radius  is proportional to the inverse of the absolute value of the cosmological constant. If these spacetimes are expressed in a suitable coordinates, the large AdS limit (flat-space limit) of these spacetimes is well-defined and yields an asymptotically flat metric which is a solution of the Einstein gravity without  cosmological constant. Correspondingly,  one can think of the analogous operation  on the field theory side. In \cite{Bagchi:2010zz,Bagchi:2012cy}, it was proposed that the flat-space limit of the gravity theory corresponds to the ultra-relativistic limit of the boundary CFT. According to \cite{Bagchi:2010zz,Bagchi:2012cy},  asymptotically flat spacetimes are holographically  dual to the ultra-relativistic field theories which are BMS-invariant and we call them BMSFT . On the gravity side, BMS symmetry is the asymptotic symmetry of the asymptotically flat spacetimes at null infinity \cite{BMS}-\cite{aspects}. On the field theory side BMS algebra is given by Inonu-Wigner contraction of the conformal algebra \cite{Bagchi:2012cy}. Thus the situation is  similar to the AdS/CFT correspondence i.e.  the asymptotic symmetry of the (d+1)-dimensional asymptotically flat spacetimes is the same as exact symmetry of the dual field theory. This duality is known as Flat/BMSFT correspondence.
 
In the context of Flat/BMSFT correspondence, one can study holographic description of BMSFT observables. An interesting non-local observable in field theory, with a well known dual gravity description, is the  entanglement entropy. In fact,  For a given sub-system $A$ with its complement $B$, the entanglement entropy measures  how much entanglement exists between the two sub-systems. Computing entanglement entropy for a generic field theory, is by no means an easy task . Nevertheless, it is possible to find universal formula for  the field theories with infinite-dimensional symmetries such as two dimensional conformal field theories (CFT$_2$)  \cite{Cardy1,Cardy2,Huerta}.

For two sub-systems A and B , it is more natural to compute the amount of correlations (both classical and quantum) between these two sub-systems which is given by the mutual information.    In fact, it is a finite quantity which measures the amount of information that A and B can share \cite{Hayden}. Subadditivity property of the entanglement entropy gurantees that mutual information is always non-negative \cite{Tonni}. Another interesting quantity to consider in this context is the 3-partite information which is defined for three disjoint sub-systems of a field theory and measures the degree of extensivity of the mutual information. Similarly, it is a finite quantity and for the field theories with holographic dual is non-positive \cite{Hayden}.

Similar to CFT$_2$, BMSFT$_2$ and BMSFT$_3$  are field theories with infinite dimensional symmetry. This may imply  that the entanglement entropy (at least for simple intervals) could have a universal form. Study of the BMSFT entanglement entropy has been started in \cite{Bagchi:2014iea} and continued in \cite{Hosseini:2015uba}-\cite{Fareghbal:2017ujy}.  An interesting holographic description for the BMSFT entanglement entropy has been introduced in \cite{Jiang:2017ecm}. The idea is very similar to that of CFTs in the context of AdS/CFT correspondence  \cite{Takayanagi}-\cite{T}. It was proposed in \cite{Jiang:2017ecm} that BMSFT entanglement entropy is given by the length  of a spacelike geodesic in the bulk which is connected to the null infinity  by  two null geodesics. 

 In this paper we use the proposal in \cite{Jiang:2017ecm} to calculate holographically the mutual information and the 3-partite information of  BMSFT$_2$. 
We  show that  these two quantities , indeed , have desired properties expected for a field theory with holographic dual. We   demonstrate that there is an interesting bound in which the mutual information takes a transition from positive value to zero  known as $"disentangling\,\, transition"$. Finally, we find that the 3-partite information takes non-positive values which is consistent with \cite{Hayden}. 

The  paper is organized as follows: In Section $2$ we introduce the Flat/BMSFT correspondence. In Section $3$  we briefly review the proposal of \cite{Jiang:2017ecm}  to study the holographic entanglement entropy  of BMSFT . Section $4$ is devoted to the holographic calculation of BMSFT$_2$ mutual information. Section $5$ contains computation of BMSFT$_2$ 3-partite information  by using flat-space holography. Finally Section $6$ is designated for conclusions and discussions.

 \section{Flat/BMSFT correspondence}
The asymptotic symmetry group (ASG) at null infinity of the asymptotically  flat spacetimes is the infinite dimensional BMS$_{3}$ group whose corresponding algebra is given by
\begin{eqnarray}\label{BMS_3}
\left[L_{m},L_{n}\right] &=& (m-n)L_{m+n} +{c_{L}\over 12} m(m^2-1) \delta_{m+n,0}~,\nonumber \\
\left[L_{m},M_{n}\right] &=& (m-n)M_{m+n}+{c_{M}\over 12}m(m^2-1)\delta_{m+n,0}~,\nonumber\\
\left[M_{m},M_{n}\right] &=& 0,
\end{eqnarray}
where $m$ and $n$ are integers and the global part  $\{L_0,L_{\pm 1}, M_0, M_{\pm 1}\}$ are the generators of Poincare symmetry. The generators $L_m$ and $M_m$ are given by taking the flat-space limit of the generators of the asymptotic symmetry of the asymptotically AdS spacetimes \cite{Barnich:2012aw,Fareghbal:2013ifa}. The flat-space limit or, the zero cosmological constant  limit, of the gravity theory is performed by taking the infinite radius limit of the asymptotically  AdS metric. 

It was proposed in \cite{Bagchi:2010zz,Bagchi:2012cy} that the holographic dual of the asymptotically flat spacetimes is a one-dimension lower BMS-invariant field theory (BMSFT). This proposal is based on the observation that the result of Inonu-Wigner contraction of the conformal algebra in two dimensions is isomorphic to \eqref{BMS_3}. Generators of the conformal symmetry in a two dimensional CFT on the plane are given by 
 \begin{equation} 
 \mathcal{L}_{n} = -e^{in\omega}\partial_{\omega},\qquad\bar{\mathcal{L}}_{n} = -e^{in\bar{\omega}}\partial_{\bar{\omega}},
 \end{equation}
 where $\omega = t+x$ , $\bar{\omega} = t-x$ and $\{t,x\}$ are  spacetime coordinates . If one starts from  the following two dimensional conformal algebra,
 \begin{eqnarray}\label{2conformal algebra}
 \left[{\mathcal{L}}_{m},{\mathcal{L}}_{n}\right] &=& (m-n){\mathcal{L}}_{m+n} +{c\over 12} m(m^2-1) \delta_{m+n,0}~,\nonumber \\
\left[\bar{\mathcal{L}}_{m},\bar{\mathcal{L}}_{n}\right] &=& (m-n)\bar{\mathcal{L}}_{m+n}+{\bar c\over 12}m(m^2-1)\delta_{m+n,0}~,\nonumber\\
\left[\mathcal{L}_{m},\bar{\mathcal{L}}_{n}\right] &=& 0,
\end{eqnarray}
 and defines the  linear combinations
 \begin{equation}
 L_{n} = \mathcal{L}_{n}-\bar{\mathcal{L}}_{-n},\qquad M_{n} =\epsilon (\mathcal{L}_{n}+\bar{\mathcal{L}}_{-n}),
 \end{equation}
 and scales coordinates as
 \begin{equation}
 t\rightarrow \epsilon t,\qquad x\rightarrow x,
 \end{equation}
  where $\epsilon$ is a constant, then  it is not difficult to check that by taking  $\epsilon\to 0$ limit , the BMS$_{3}$ algebra \eqref{BMS_3} will be generated   \cite{Bagchi:2012cy}. 
 The central charges of the conformal  and BMS algebra are related  by  $c_{L} = {c-\bar{c}}$ and $c_{M} = {\epsilon(c+\bar{c})}$.

 The correspondence between asymptotically flat spacetimes and BMSFTs is known as    \,\,\, Flat/BMSFT. Using this duality one can find the universal properties of BMSFTs by merely performing calculations on the  gravity side (see \cite{Prohazka:2017lqb} for a complete list of related papers). In the rest of this paper we  use the above mentioned duality to study the mutual information and 3-partite  information of BMSFTs.

   \section{BMSFT entanglement entropy using flat-space holography}
   In order to perform  the holographic calculation of the mutual and 3-partite information, we need to introduce the holographic entanglement entropy of  BMSFT. In this section, we review the holographic entanglement entropy of CFT and BMSFT  in the context of AdS/CFT and Flat/BMSFT correspondences.
  
For an arbitrary quantum field theory in d-dimensions, there are specific degrees of freedom associated with any spatial regions. If we  decompose the total system into two sub-systems $ A $ and $ B $, the total Hilbert space $ \textit{H} $ becomes a direct products,
\begin{eqnarray}
\textit{H} =\textit{H}_{A} \otimes \textit{H}_{B}.
\end{eqnarray}
For a given decomposition, one can ask how the degrees of freedom in the region $ A $ are entangled with those of the region $ B $. One simple quantitative measure of this entanglement is the entanglement entropy.  The reduced density matrix $\rho _{A} $, for the sub-system $ A $, is given by tracing out the whole system density matrix, $ \rho $,  with respect to $ \textit{H}_{B} $,
\begin{eqnarray}
\rho _{A}=\mbox{Tr}_{B} [\rho].
\end{eqnarray}
 Then the entanglement entropy is defined as the Von-Neumann entropy of $\rho _{A} $,
\begin{eqnarray}
S _{A}=-\mathrm{Tr} [ \rho _{A} \log{\rho _{A}} ].
\end{eqnarray}
A holographic description of the entanglement entropy  has been considered  in  the context of AdS/CFT in \cite{Takayanagi}. It is proposed that  the entanglement entropy $ S_{A} $ in a CFT  can be holographically calculated by the $\text{RT} $ formula \cite{Takayanagi,Ryo,T},
\begin{eqnarray}\label{RT formukla}
S_{A}=\mathrm{ext}_{\Sigma_{A}} \left[ \frac{Area (\Sigma_{A})}{4 G_{N}}\right],
\end{eqnarray}
where $\Sigma_{A}$ is a co-dimension two surface which satisfies $ \partial \Sigma_{A} = \partial A $ and  is homologous to the region $A$ in the boundary. $G_{N}$  is the Newton constant of the  bulk theory. Therefore one can  extremize the area of $\Sigma_{A}$ to calculate the entanglement entropy.

\begin{figure}[h] \label{fig:Pen1}
   \centering
    \includegraphics[width=0.65\textwidth]{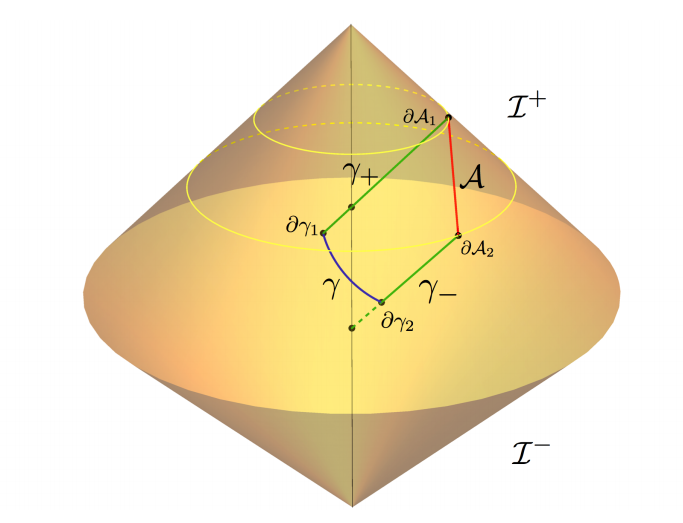} 
 \caption{The boundary interval is the red line $\mathcal{A}$, living  on the future null infinity $\mathcal{I}^+$, $\partial \mathcal{A}_{1}$ and $\partial \mathcal{A}_{2}$ are the endpoints of  interval on the boundary.
 The blue line $\gamma$ is a spacelike geodesic. The two green lines $\gamma_+$, $\gamma_-$ are null geodesics, then the entanglement entropy is given by $S_{\text{HEE}}= {\text{Length}  (\gamma) \over 4G}={\text{Length}  (\gamma_{\mathcal A}) \over 4G}$ \cite{Jiang:2017ecm}.
} 
\end{figure}

In \cite{Jiang:2017ecm}  the authors provide a nice holographic description of the BMSFT entanglement entropy. They pointed out that the holographic entanglement entropy is given by the length of a spacelike geodesic $\gamma$ in the asymptotically flat spacetime which is connected to  the   null infinity  by two null geodesics $\gamma_{\pm}$ (Fig.~ \ref{fig:Pen1}). The holographic entanglement entropy formula in this case is similar to the RT-formula \eqref{RT formukla} and is given by
\begin{eqnarray}
S_{HEE} =\frac{L_{\gamma}}{4G}.
\end{eqnarray}
  If one characterizes the sub-system A in  the BMSFT by
\begin{eqnarray}
A_{reg} :\,\,\,\,\,\,(-\frac{l_{u}}{2} +\epsilon _{u} ,-\frac{l_{\phi}}{2} +\epsilon _{\phi}) \rightarrow (\frac{l_{u}}{2} -\epsilon _{u} ,\frac{l_{\phi}}{2} -\epsilon _{\phi}),
\end{eqnarray}
where $\epsilon _{u} $ and $\epsilon _{\phi} $ are  cut-offs to regulate the interval, then entanglement entropy of the above interval  for the zero temperature BMSFT on the plane  and finite temperature BMSFT on the cylinder  are, respectively, given by
\begin{equation} \label{zerotem}
S_{EE}(l_{u},l_{\phi}) =\frac{c_{L}}{6} \log\frac{l_{\phi}}{\epsilon_{\phi}}+ \frac{c_{M}}{6} (\frac{l_{u}}{l_{\phi}}-\frac{\epsilon _{u}}{\epsilon _{\phi}}),
\end{equation}
\begin{eqnarray}\label{finite}
 S_{EE}(l_{u},l_{\phi})=\frac{c_{L}}{6} \log(\frac{\beta _{\phi}}{\pi \epsilon _{\phi}}\,\, \sinh(\frac{\pi l_{\phi}}{\beta _{\phi}}))   +\frac{c_{M}}{6 \beta _{\phi}} [\pi (l_{u} +\frac{\beta _{u} l_{\phi}}{\beta _{\phi}})\,\, \coth(\frac{\pi l_{\phi}}{\beta _{\phi}}) -\beta _{u}]  -\frac{c_{M} \epsilon_{u}}{6\epsilon _{\phi}}, 
\end{eqnarray}
where $\beta _{u} $ and $\beta _{\phi} $ are thermal identifications of the coordinates in the thermal BMSFT \footnote{We follow  convention of \cite{Jiang:2017ecm} in which $\beta _{u} $ and $\beta _{\phi} $ are negative quantities. }. 
  In this paper we consider the BMSFTs dual to the Einstein gravity where  $c_{L}$ is zero \cite{Barnich:2006av}.
\\

\section{BMSFT Mutual information and its holographic calculation }
In this section  we start from definition of the mutual information in any field theory and then  calculate BMSFT mutual information holographically. 
\subsection{Definition of mutual information }
In a quantum field theory, entanglement entropy of a region $A$ contains short-distance or high energy divergence. In fact, in an unregulated quantum field theory the entanglement entropy is formally divergent due to the presence of high energy singularities associated with the boundary law  behaviour. However, there is a quantity, called the mutual information which is an appropriate linear combination of the entanglement entropy and remains finite in a quantum field theory.  The mutual information of  two  sub-systems $A$ and $B$  is defined by,
\begin{eqnarray}
I(A,B)=S_{A}+S_{B}-S_{A\cup B}\label{MI}
\end{eqnarray}
where $S_{X}$ denotes the entanglement entropy of the region $X$ . Mutual information  measures the total correlations between the two sub-systems $A$ and $B$. Furthermore, 
 it is positive semi-definite quantity that is proportional to the entanglement entropy when $B\equiv A^{c}$,  where $A^{c}$ indicates the complement of $A$, such that $S_{A\cup A^{c}}=0$.
It was shown in \cite{Headrick:2010zt} that in the holographic dual theories, mutual information indeed undergoes a " first order phase transition " as the separation between the two sub-systems $A$ and $B$ is increased. In other words, for small separation $I(A,B)\neq 0$, but for large separation $I(A,B)= 0$. When $I(A,B)= 0$, the two sub-systems $A$ and $B$ become completely decoupled hence one would call it a " disentangling transition ". Furthermore, if  $A$ and $B$ together  cover the entire system then clearly $S_{A\cup B}=0$ and $I(A,B)=2S_{A}=2S_{B}$ \cite{Kundu}.

\subsection{Holographic BMSFT Mutual information in zero temperature}
\begin{figure}[h]
   \centering
    \includegraphics[width=0.65\textwidth]{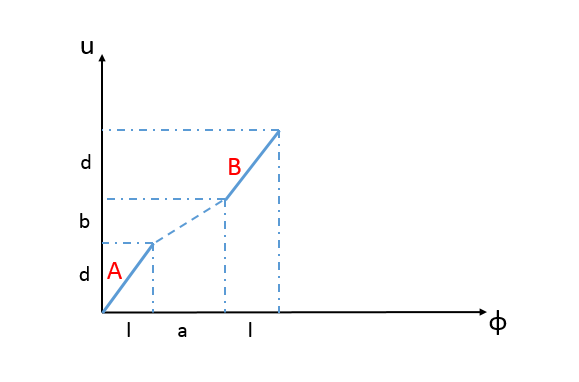} 
 \caption{Two disjoint entangled regions for calculating mutual information}
 \label{fig:two}
 \end{figure}
 In the following we consider a two dimensional  BMSFT  living on a plane whose coordinates are   $(u,\phi)$.  Sub-systems $A,B$ are two intervals  depicted in (Fig.~\ref{fig:two}).
 
 \begin{figure}[h]
\centering
\includegraphics[width=0.95\textwidth]{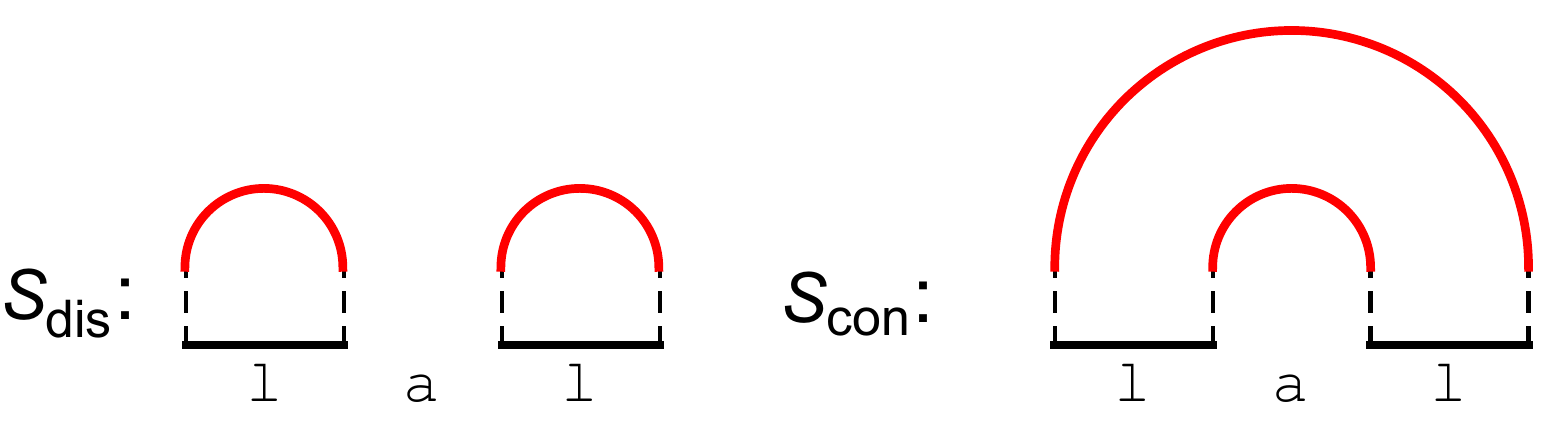} 
\caption{Two different configurations for computing $S_{A\cup B}$.The time coordinate is suppressed.}
\label{fig:MI}
\end{figure}

For a  single interval, the zero temperature  entanglement entropy is given by \eqref{zerotem}. Thus $S_{A}$ and $S_{B}$ in the mutual information formula \eqref{MI} can be easily computed.   There are two different candidate configurations  for  calculation of  the entanglement entropy of the union region $S_{A\cup B}$ (Fig. \ref{fig:MI}). Similar to the holographic mutual information in the context of AdS/CFT \cite{Tonni,Kundu,T1,Ali}, $S_{A\cup B}$ is given by one of these configurations which has minimal length. Depending on the length of intervals and their separation we have
 
\begin{eqnarray}\label{sAB}
		{S_{A\cup B}}=
		\begin{cases}
			S_{dis}=2 S(d,l) & \frac{d}{l}< \frac{b}{a},  \\
			S_{con}=S(2d+b,2l+a) +S(b,a) & \frac{d}{l}  >\frac{b}{a}.  \\
		\end{cases}
	\end{eqnarray}

Thus, there  is a critical point of parameters at which the minimum configuration is transitioned from the disconnected configuration to the connected one.  Consequently, using \eqref{zerotem}, \eqref{MI} and \eqref{sAB}, for the two disjoint entangling regions depicted in (Fig. \ref{fig:two}), the holographic mutual information  becomes,
\begin{eqnarray}\label{MIZ}
		{I(A,B)}=
		\begin{cases}
			0 \qquad\frac{d}{l} <\frac{b}{a}  \\
			2S(d,l) - S(2d+b,2l+a) - S(b,a) \qquad\frac{d}{l}  >\frac{b}{a}  \\
		\end{cases}
	\end{eqnarray}
The most significant point is that there  is a bound  for the choices of two sub-systems  to have entanglement correlation. Furthermore, it can be easily shown that $I(A,B)$ is positive for  $\frac{d}{l}>\frac{b}{a}$. When $I(A,B)=0$   the  two sub-systems A and B    become completely decoupled hence one can say that a disentangling transition occurs. Interestingly, according to (Fig. \ref{fig:two}), there is a  geometric interpretation  of the transition  point i.e. $\frac{d}{l}=\frac{b}{a}$ which indicates that the intervals and their separation should be  along a line. As a result,  the  intervals and their separation angles with   $\phi$-coordinate     indeed determine the amount of correlation. Thus two large intervals with very small separation can be entangled or disentangled depending on their angles. This strange result is a consequence of  extensions of intervals in the $u$-coordinate. We believe that the ultra-relativistic aspect of BMSFTs may justify this observation.

\subsection{Holographic BMSFT Mutual information in finite temperature}
In this subsection we calculate the mutual information of two disjoint intervals (Fig.~\ref{fig:two}) of finite temperature BMFST . Using (\ref{finite}), $S_{A}$  and $S_{B}$ are easily obtained. Analogously, to compute the entanglement entropy $S_{A\cup B}$, there are two possible configurations   (Fig.~\ref{fig:MI})  which contribute to the mutual information in different ranges of parameters.    Defining   $X\equiv\frac{\pi a}{\beta_{\phi}}$, $Y\equiv\frac{\pi l}{\beta_{\phi}}$, $W\equiv\frac{\pi b}{\beta_{\phi}}$, $Z\equiv\frac{\pi d}{\beta_{\phi}}$ and $\alpha\equiv\frac{\beta_{\phi}}{\beta_{u}}$ the entanglement entropy  $S_{A\cup B}$  is given  by
\begin{eqnarray}\label{sab finite}
		{S_{A\cup B}}=
		\begin{cases}
			2 S(d,l) & \gamma > T,  \\
			S(2d+b,2l+a) +S(b,a) & \gamma < T,   \\
		\end{cases}
	\end{eqnarray}
where	
\begin{equation}
	\gamma=\frac{X+\alpha W}{2(Y+\alpha Z)},\qquad T=\frac{\coth(Y) - \coth(X+2Y)}{\coth(X)+\coth(X+2Y)}.
	\end{equation}

Using \eqref{sab finite} and  definition of the mutual information \eqref{MI}, we find the holographic mutual information as,
\begin{eqnarray}\label{MIfinite}
		{I(A,B)}=
		\begin{cases}
			0 &\gamma > T,  \\
			2S(d,l) - S(2d+b,2l+a) - S(b,a) & \gamma < T.  \\
		\end{cases}
	\end{eqnarray}

It is an easy task to show that $I(A,B)$ is positive for $\gamma < T $. Similarly, one can clearly observe the transition of the mutual information from positive values to zero in finite temperature and hence an entangling transition occurs. Consequently, one can claim that BMSFTs in both zero  and finite temperature regime  respect the subadditivity condition \cite{Tonni}  
\begin{eqnarray}
S(A)+S(B)\geq S(A\cup B).
\end{eqnarray}

The mutual information of the zero temperature BMSFT on the cylinder is obtained by using \eqref{MIfinite}  if one substitutes  $\beta_{u}=0$ and $\beta_{\phi}=-2\pi i$. In this case we have
\begin{eqnarray}\label{cylinder}
		{I(A,B)}=
		\begin{cases}
			0 &\frac{b}{d}>\frac{\sin\frac{a}{2}}{\cos\frac{(a+l)}{2}\sin\frac{l}{2}},  \\
			2S(d,l) - S(2d+b,2l+a) - S(b,a) & \frac{b}{d}<\frac{\sin\frac{a}{2}}{\cos\frac{(a+l)}{2}\sin\frac{l}{2}}.  \\
		\end{cases}
	\end{eqnarray}
In  the  limit $l , a \rightarrow 0$, \eqref{cylinder} is changed to \eqref{MIZ} which can be considered as a consistency check of our calculation.
\section{BMSFT 3-partite information and its holographic calculation }

Another useful and interesting quantity that can be defined by using  the entanglement entropy, is the 3-partite information,
\begin{equation} \label{tri}
 I_{3}(A,B,C)\equiv S_{A}+S_{B}+S_{C}-S_{A\cup B}-S_{A\cup C}-S_{B\cup C}+S_{A\cup B\cup C},
\end{equation}
 where $A$, $B$ and $C$ are three disjoint regions and $S_{A\cup B\cup C}$ is the entanglement entropy for the union of three sub-systems. Similar to the mutual information, 3-partite information is free of divergences and finite. This quantity can also be positive, negative or zero depending on the underlying field theory \cite{Casini:2008wt}. However, it has been shown that for a field theory with a holographic dual the 3-partite information is always non-positive, i.e. $I_{3}(A,B,C)\leq0 $  \cite{Hayden}. $I_{3}$  is a measure of extensivity of the mutual information; in fact, it can be written in terms of the mutual information as
 \begin{eqnarray}
I_{3}(A,B,C)\equiv I(A, B) + I(A, C) - I(A, B\cup C).
 \end{eqnarray}
  Accordingly, the mutual information is subextensive when $I_{3}>0$, extensive when $I_{3}=0$ and superextensive when $I_{3}<0$. In either the extensive or the superextensive case the mutual information is said to be monogamous. 

3-partite information of the sub-systems in the field theories which have holographic dual can be calculated by using holographic methods. 
 In \cite{Tonni} the authors considered  quantum systems whose gravity duals are Vaidya spacetimes in three and four dimensions.
 They showed that when the null energy condition is violated the holographic 3-partite information takes positive values  for specific ranges of time. As a result, the holographic mutual information becomes non monogamous. In other words, they find that the null energy condition is a necessary condition both for the strong subadditivity of the holographic entanglement entropy and for the monogamy of the holographic mutual information.
 
 In the rest of this paper we use Flat/BMSFT correspondence to calculate the BMSFT 3-partite information.  Among the terms occurring in the definition of the holographic 3-partite information, \eqref{tri},  computation of $S_{A\cup B\cup C}$ is more challenging.  
 
\begin{figure}[h]
   \centering
    \includegraphics[width=0.65\textwidth]{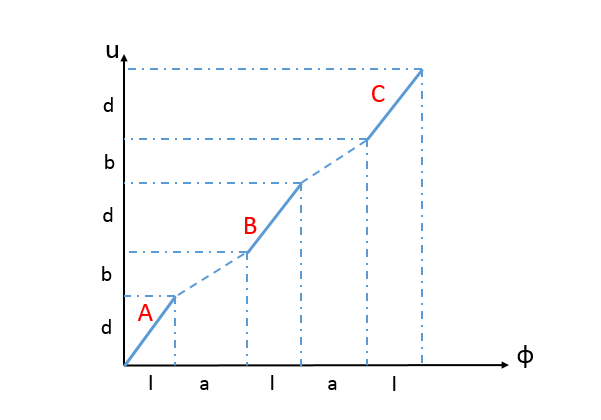} 
 \caption{Three disjoint entangling regions for calculating 3-partite information}
 \label{fig:tri}
 \end{figure}

\begin{figure}
               \includegraphics[width=0.99\textwidth]{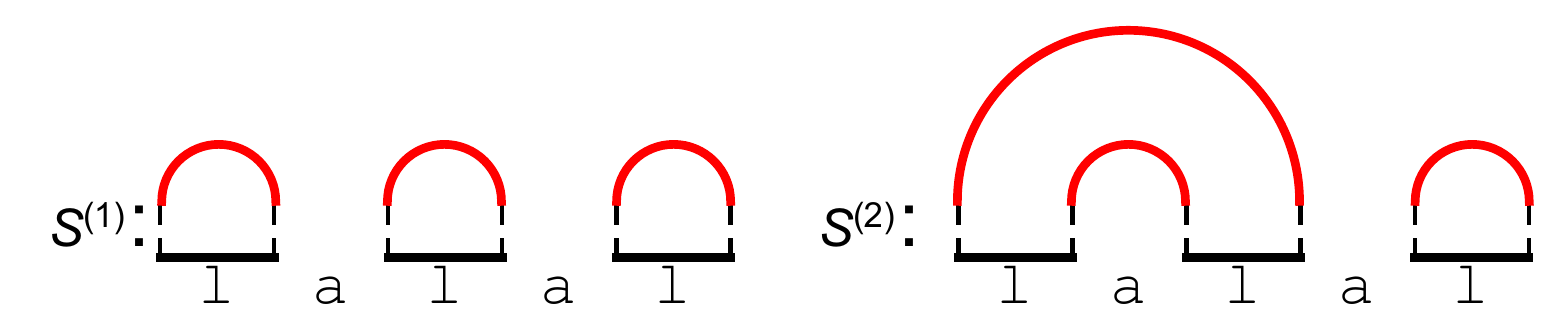}  
           
        \includegraphics[width=0.99\textwidth]{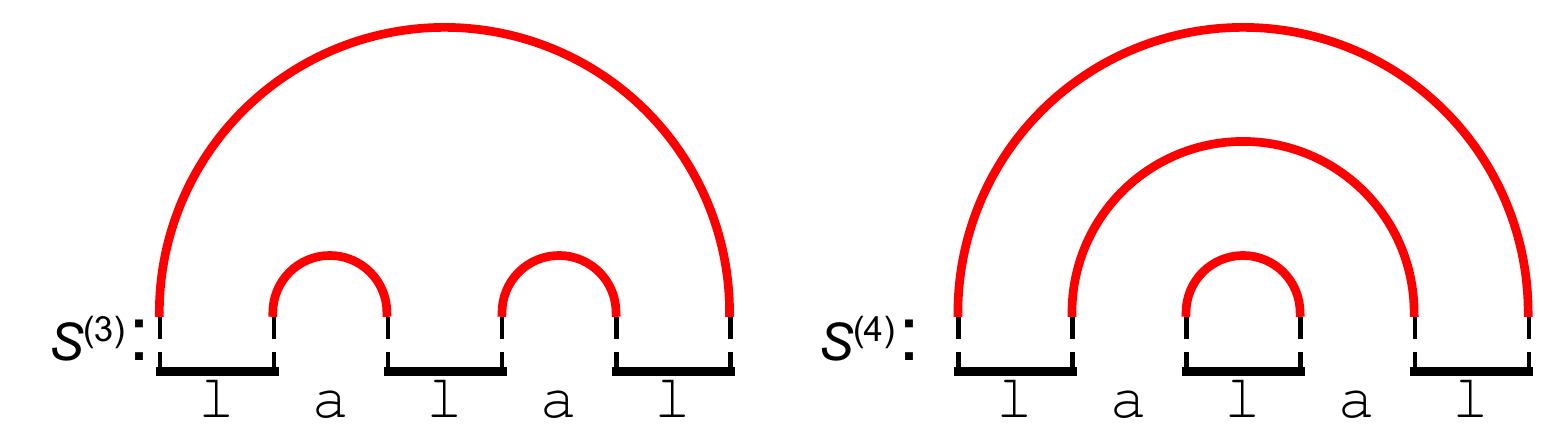} 
       
  \caption{Schematic configurations of surfaces to consider in the computation of $S_{A\cup B\cup C}$ .The time coordinate is suppressed.}
\label{fig:fourconf}
\end{figure}

Here, we consider a $2-$dimensional  BMSFT living on a plane whose coordinates are denoted by  $(u,\phi)$. The three disjoint intervals   $A,B,C$ are depicted in (fig.  \ref{fig:tri}). In order to calculate the holographic 3-partite information of  these sub-systems, it is necessary to   compute  $S_{A\cup B\cup C}$ at first stage.
 In principle, for $N$ entangling regions (or $N$ intervals ) one should compare $(2N-1)!!$ configurations, which is $15$ in our case $(N=3)$. However, it has been shown that for $N=3$ we are left only with the four independent candidates  which are schematically shown in (Fig.~\ref{fig:fourconf}) \cite{Tonni}. Thus , $S_{A\cup B\cup C}$ is given  by the minimum area  of the underlying configurations.  
     
If we consider zero temperature BMSFT, using \eqref{zerotem}, we find the following expressions for the union of two and three intervals
\begin{align}
{S_{A\cup B}}={S_{B\cup C}}=
		&\begin{cases}
			2 S(d,l) &\frac{d}{l} <\frac{b}{a},  \\
			S(2d+b,2l+a) +S(b,a) &\frac{d}{l}  >\frac{b}{a},  \\
		\end{cases}\\
	{S_{A\cup C}}=
		&\begin{cases}
			2 S(d,l) &\frac{d}{l} <\frac{b}{a},  \\
			S(3d+2b,3l+2a) +S(d+2b,l+2a) &\frac{d}{l}  >\frac{b}{a},  \\
		\end{cases}\\
	{S_{A\cup B\cup C}}=
		&\begin{cases}
			3 S(d,l) &\frac{d}{l} <\frac{b}{a},  \\
			S(3d+2b,3l+2a) +2S(b,a) &\frac{d}{l}  >\frac{b}{a}.  
		\end{cases}
\end{align}	
Substituting these results into \eqref{tri} the holographic 3-partite information of the zero temperature BMSFT reads,
\begin{align}\label{3p zero}
{I_{3}(A,B,C)}=
		\begin{cases}
			0 &\frac{d}{l} <\frac{b}{a},  \\
			3S(d,l) - 2S(2d+b,2l+a) - S(d+2b,l+2a) & \frac{d}{l}  >\frac{b}{a}.   
		\end{cases}
\end{align}
Using \eqref{zerotem} and \eqref{3p zero}, it is not difficult  to show that the 3-partite information is always negative for  $\frac{d}{l}>\frac{b}{a}$ . Consequently, the holographic mutual information of the zero temperature BMSFT becomes monogamous which is consistent with \cite{Hayden}.

The main subtlety to calculate the  entanglement entropy of union of sub-systems also appears in the computation of 3-partite information of the finite temperature BMSFT.  In order to compute  $S_{A\cup B\cup C}$, we need to use \eqref{finite} to find the minimal surface among the configurations in (fig.\;\ref{fig:fourconf}). To obtain clear analytic results, we have to do this calculation in  particular limits.

 In the  limit $a,l<<1$,  we obtain 
\begin{align}\label{3p-finite small}
{I_{3}(A,B,C)}=
		\begin{cases}
			0 &\frac{d}{l} <\frac{b}{a},   \\
			3S(d,l) - 2S(2d+b,2l+a) - S(d+2b,l+2a) & \frac{d}{l}  >\frac{b}{a},   
		\end{cases}
\end{align}
In the latter case, it is straight forward to show that  $I_{3}(A,B,C)<0$. Similarly,  the transition  point i.e. $\frac{d}{l}=\frac{b}{a}$ has the same geometric description as \eqref{MIZ} which states that the holographic mutual information becomes monogamous  if the intervals and their separation lie  along a line in the $(u,\phi)$ plane. 
On the other hand, in the regime $a,l>>1$, we get 
\begin{align}
{I_{3}(A,B,C)}=0.
\end{align}

Consequently, the 3-partite information of the finite temperature BMSFT is non-positive in both very large and very small intervals $a,l$. Since expression of $I_{3}(A,B,C)$ is continuous between these two limits,  it increases from  negative values  in $a,l<<1$ to  zero  in  $a,l>>1$.  Subsequently, the mutual information  of the finite temperature BMSFT   is also monogamous
\section{Conclusion}
In this paper, using flat-space holography, we studied the holographic mutual information of a two dimensional BMSFT which is dual to three dimensional  asymptotically flat spacetimes. We found that , in both zero and finite temperature regimes, the mutual information does respect the strong subadditivity inequality which states that $I(A,B)\geq0$.  In other words, a disentangling transition occurs as two sub-systems become decoupled.  Furthermore,  there is a bound for the choices of sub-systems of  BMSFT above which there is non-vanishing correlation between the two sub-systems. Considering the holographic 3-partite information, we observed that the holographic mutual information is monogamous $i.e.$ $I_{3}(A,B,C)\leq0$.

The appearance of both the disentangling transition and the monogamous mutual information  are common and important properties   which  one expects in holographic theories. In this sense, BMSFTs as the dual of asymptotically flat spacetimes  are not very strange theories. However, in order  to get non-zero mutual and 3-partite information the intervals must be extended in the time coordinate. Since  BMSFTs are ultra-relativistic theories  dividing intervals into spacelike, timelike and null  dose not have clear meaning and this fact should be considered to justify the unusual resultant bounds.  The uncommon increase  or decrease  of BMSFT n-partite information might have its roots in the time-dependent intervals. The consequences of this weird behaviour of n-partite information is an interesting subject for the future works. 
\subsubsection*{Acknowledgements}
The authors would like to thank  Seyed Morteza Hosseini and Pedram Karimi for useful comments.
\appendix


\end{document}